\documentclass{pos}

\usepackage{amsmath}
\usepackage[numbers,sort&compress]{natbib}
\newcommand{\preprint}{
  \begin{picture}(0,0)
    \put(-55,100){{\rm\normalsize ADP-08-16/T676}}
  \end{picture}}

\title{The strong-coupling limit of lattice Landau
  gauge\footnotetext{Supported by the Australian Research Council and
    by eResearch South Australia.}\preprint}

\ShortTitle{The strong-coupling limit of lattice Landau gauge}

\author{\speaker{Andr\'e Sternbeck} and  
        Lorenz von Smekal\\
        Centre for the Subatomic Structure of Matter
        (CSSM), School of Chemistry \& Physics,\\
        The University of Adelaide, SA 5005, Australia\\
        \mbox{E-mail: \email{andre.sternbeck@adelaide.edu.au},
        \email{lorenz.smekal@adelaide.edu.au}}
}

\abstract{We report on our recent study of the gluon and ghost
  propagators of pure SU(2) minimal lattice Landau gauge in the
  strong-coupling limit. In this limit, we find evidence of the
  conformal infrared behaviour of these propagators as predicted by
  functional continuum methods. However, in the strong-coupling limit
  this happens for lattice momenta with $a^2q^2>1$, in units of the
  lattice spacing $a$. Deviations from conformal scaling for 
  $a^2q^2<1$ are well parameterised by a transverse gluon mass. A
  comparison of various lattice definitions of gauge potentials, all
  equivalent in the continuum limit, shows that (a) both the critical
  exponent and coupling can be extracted unambiguously from the
  high-momentum data in the strong-coupling limit, in good agreement
  with the continuum predictions; but that on the other hand (b) the
  massive branch depends on the definition of lattice gluon fields and
  is thus not unambiguously defined. We demonstrate that this
  ambiguity is also present in the low-momentum region for commonly
  used values of the lattice coupling in SU(2).}

\FullConference{8th Conference Quark Confinement and the Hadron Spectrum \\
		 September 1--6, 2008\\
		 Mainz, Germany}

\newcommand{\Fig}[1]{Fig.~\ref{#1}}

\newcommand{\identity}{\mathbf{1}}

\begin{document}

\section{Introduction}

Continuum functional methods favour QCD's gluon and ghost propagators 
in Landau-gauge to show a conformal infrared behaviour where their
respective dressing functions behave as
\cite{vonSmekal:1997isvonSmekal:1997vx,Lerche:2002ep,
  Zwanziger:2001kw,Pawlowski:2003hq}
\begin{equation} \label{eq:infrared-gh_gl}
  Z(p^2) \sim (p^2/\Lambda^2_{\mathrm{QCD}})^{2\kappa_Z}\;, 
\qquad
  G(p^2) \sim (p^2/\Lambda_\mathrm{QCD}^2)^{-\kappa_G}
\qquad
\text{for}\quad p^2 \to 0\;, 
\end{equation}
which are both determined by an unique critical
infrared exponent $\kappa_Z = \kappa_G \equiv \kappa$ with $ 0.5 <
\kappa < 1$. Under a mild regularity assumption on the ghost-gluon
vertex \cite{Lerche:2002ep}, the value of this exponent is furthermore
obtained as $\kappa \approx 0.595$
\cite{Lerche:2002ep,Zwanziger:2001kw}. The conformal nature of this
infrared behaviour in the pure 
Yang-Mills sector of Landau gauge QCD is evident in the generalisation
to arbitrary gluonic correlations \cite{Alkofer:2004it}. In particular,
in the limit $p\to0$ the ghost-gluon vertex is infrared finite, and the
non-perturbative running coupling in Eq.~(1.2) 
~\cite{vonSmekal:1997isvonSmekal:1997vx}
approaches an infrared fixed-point, $\alpha_s \to \alpha_c$
\begin{floatingfigure}[r]
\begin{minipage}[b]{5.6cm}\vspace{-2ex}
  \begin{equation} 
    \label{alpha_minimom}
    \alpha_s(p^2) \, = \, \frac{g^2}{4\pi} Z(p^2) G^2(p^2)
  \end{equation}
\end{minipage}
\end{floatingfigure}
\noindent 
whose maximum value is $\alpha_c \approx 4.46$ for SU(2)
\cite{Lerche:2002ep}. It has been shown that in presence of the single
scale, $\Lambda_{\mathrm{QCD}}$, the solution with such an infrared 
behaviour is unique \cite{Fischer:2006vf}. 

To observe this \emph{scaling solution}, at least in an approximation,
in lattice simulations in a finite box of extend $L$, a wide
separation of scales, $\pi/L \, \ll p \, \ll \Lambda_{\mathrm{QCD}}$,
is necessary such that a reasonably large number of modes with momenta
$p$ sufficiently far below $\Lambda_\mathrm{QCD}$ are accessible whose
corresponding wavelengths are at the same time much shorter than
$L$. Despite tremendous efforts
\cite{Sternbeck:2007ug,Bogolubsky:2007ud,Cucchieri:2007md}
the majority of lattice investigations, however, could not confirm this scaling 
solution. Rather, compelling agreement between lattice Landau gauge
and continuum results has been found when the restriction to the first
Gribov region is implemented. In such a case gluons and ghosts
decouple at low momenta, due to the appearance of a transverse gluon
mass (i.e. an infrared-finite gluon propagator) which leads to an
essentially free ghost propagator with a free massless-particle
singularity at zero momentum. This type of solution is not within the
class of scaling solutions, and it is termed the {\em
  decoupling solution} in contradistinction \cite{Fischer:2008uz}.

In \cite{Sternbeck:2008wgSternbeck:2008mv} we recently reported on
our study of the gluon and ghost propagators in the strong-coupling
limit, $\beta \to 0$, of pure $SU(2)$ lattice Landau gauge. This
unphysical limit, which can be interpreted as the formal limit
$\Lambda_\mathrm{QCD} \to \infty $, allows us to assess whether the
predicted conformal behaviour can be seen for the larger lattice
momenta $p$, after the upper bound $p \, \ll \Lambda_{\mathrm{QCD}}$
has been removed, in a range where the dynamics due to the gauge
action would otherwise dominate and cover it up
completely. Furthermore, the strong-coupling limit provides a powerful
tool to study the non-perturbative measure for gauge-orbit space in
Landau gauge. It is this measure that is being assessed when the
gauge-field dynamics is switched off. It turns out that there is a
discretisation ambiguity which manifests itself in dependencies on the
lattice definition of gauge fields underlying the respective lattice
Landau gauges and their measures. The strong-coupling limit serves to
isolate this ambiguity which noticeably affects the decoupling branch
at $a^2 q^2 < 1$. Nonetheless, it is possible to extract infrared
critical exponent and coupling at large $a^2q^2$ consistent with the
scaling solution, and unaffected by the discretisation ambiguity.

\section{Infrared exponents}

We simulate pure $SU(2)$ gauge theory in the strong-coupling limit by
generating random link configurations $\{U\}$. These are sets of
$SU(2)$ gauge links, $U_{x\mu} = u^0_{x\mu}\identity + i\sigma^a
u^a_{x\mu}$, equally distributed over~$(u^0,\vec{u})_{x\mu}\in
S^3$. Those configurations are fixed to the minimal lattice Landau
gauge and gluon and ghost propagators are then calculated in momentum
space employing standard techniques (see
\cite{Sternbeck:2008wgSternbeck:2008mv} for further details).

The gluon propagator in the strong-coupling limit is observed to
increase with momentum, while it plateaus at low momenta. This massive
behaviour sets in, irrespective of the lattice size ($N=L/a$), at
around $x\equiv a^2q^2 \approx 1$, and the observed mass behaves as   
$
  M^2 \equiv \lim_{x\to 0}   D^{-1}(x) \, \propto \, 1/a^2
$
with hardly any significant dependence on $N$.  
In particular, if there is a systematic $N$ dependence at all, 
the zero momentum limit of the gluon propagator tends to slowly
increase with the volume. It certainly 
extrapolates to a finite value $\propto 1/a^2$ in the
infinite-volume limit, $N \to \infty\,$.

\begin{floatingfigure}[r]
\parbox{5.8cm}{\includegraphics[width=6.1cm]{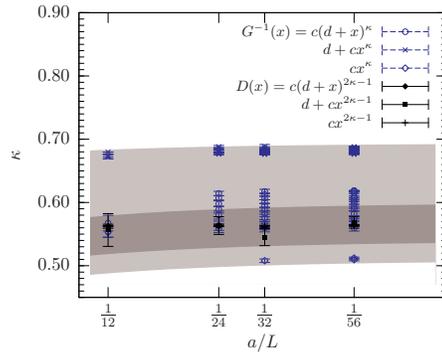}
  \vspace{-4ex}
  \caption{$\kappa$ versus $a/L$ for the ghost and gluon
    propagators. Grey-coloured bands mark the variation of $\kappa$
    with the fit model.}
  \label{fig:kappa_gl_gh_L_fits}}
\end{floatingfigure}
In order to assess the asymptotic form of the gluon dressing function
at large lattice momenta we have fitted the gluon
propagator data to different fitting formulas. The results of these fits
for the gluon exponent $\kappa_Z$ from different fit models and
lattice sizes are shown in \Fig{fig:kappa_gl_gh_L_fits}. The observed
dependencies on either one are rather small.  There is a general trend
for $\kappa_Z$ to slightly increase with $a/L$ (the dark grey band in
\Fig{fig:kappa_gl_gh_L_fits}) though this is within the systematic
uncertainty due to the fit model. 

Similar fits were performed to extract the exponent $\kappa_G$ from
the ghost dressing function $G$. These fits are less robust with a
more pronounced model dependence (light grey band in
\Fig{fig:kappa_gl_gh_L_fits}). This is mainly due to the wider
transition region, from $G = $ const.~at small $x$ to $G\sim
x^{-\kappa_G} $ at large $x$, which is under less control here.  
The exponent can nevertheless be estimated as $\kappa_G = 0.60(7)$. 
The results are consistent with the scaling relation $\kappa_Z = \kappa_G$.

\vspace{-1ex}
\section{Different gauge-field definitions on the lattice}

The strong-coupling limit is an ideal testbed for different lattice
definitions of gauge-fields which correspond to different choices of
coordinates that agree only near the identity, or in the continuum
limit. The definition of the standard lattice Landau gauge (SLG),
e.g., corresponds to choosing separate coordinates for the Northern
(NH) and Southern Hemispheres (SH) of $S^3$ in the case of
$SU(2)$. Strictly speaking, the SLG gluon propagator therefore
corresponds to an average for each link of the contributions from NH
and SH to the expectation value.
The maximal chart is provided by stereographic projection which covers
the whole sphere except for the South Pole. A definition of $SU(2)$ gauge
fields on the lattice based on stereographic projection is possible
(see \cite{Sternbeck:2008wgSternbeck:2008mv}). It agrees with the
standard definition near the North Pole, and in the continuum limit,
but the South Pole is now at infinity and the gauge fields are
non-compact. The associated Landau gauge is the modified lattice
Landau gauge (MLG) of Ref.~\cite{vonSmekal:2007ns}.

When comparing MLG to the ever popular SLG, there is no advantage that the
SLG has over the MLG. Both lattice definitions of Landau gauge have
the same continuum limit, and any differences between MLG and SLG data
at finite lattice spacings are lattice artifacts. Furthermore, their
lattice Landau gauge conditions define gluon fields that are
transverse in the physical momentum $aq_\mu=2\sin(\pi k_\mu/N_\mu)$
(with $k_\mu\in(-N_\mu/2,N_\mu]$) at any finite lattice spacing $a$.

\begin{figure*}
  \centering
  \begin{minipage}[t]{0.48\linewidth}
    \includegraphics[height=5.cm]{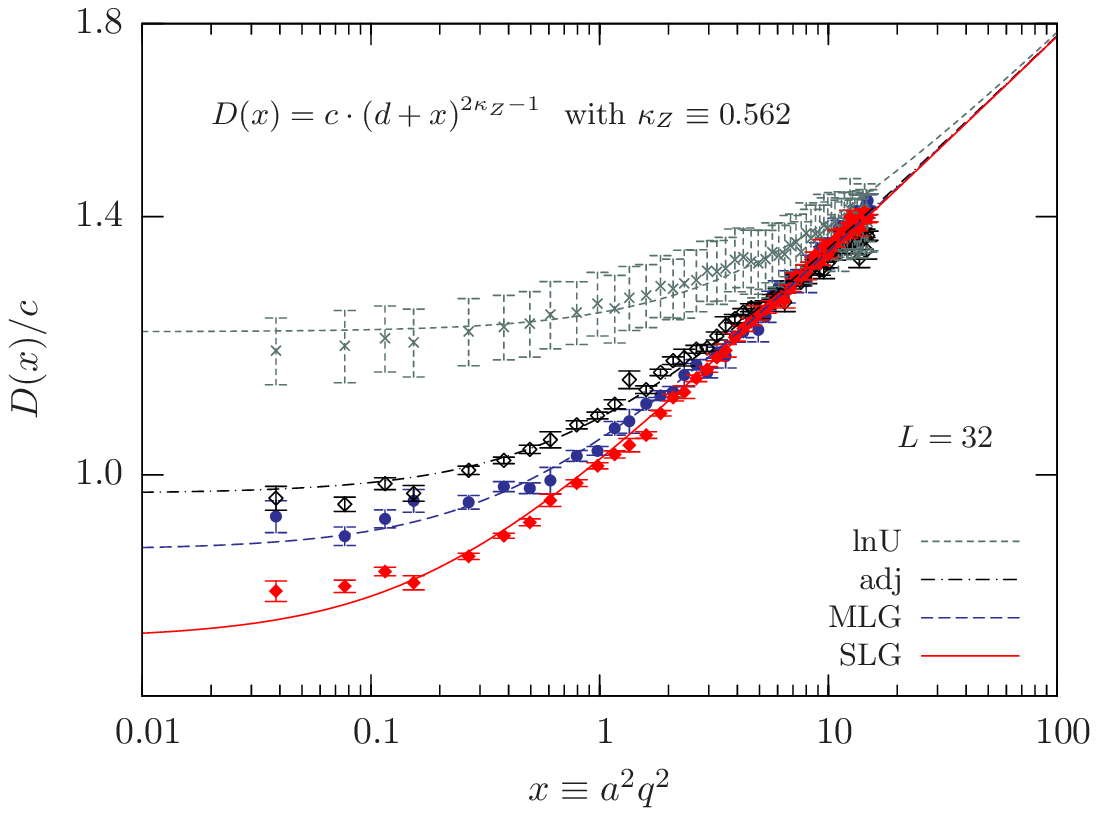}
    \caption{The strong-coupling gluon propagator over
      $a^2q^2$ for the various definitions of gauge fields.
      All on $32^4$ lattices and normalised to the 
      scaling branch after fitting to \mbox{$D(x)=c(d+x)^{2\kappa_Z-1}$};
      all with $\kappa_Z=0.562$ from the fit to the SLG data.} 
    \label{fig:gl_qq_beta0}
  \end{minipage}
  \hfill
  \begin{minipage}[t]{0.48\linewidth}
    \includegraphics[height=5.cm]{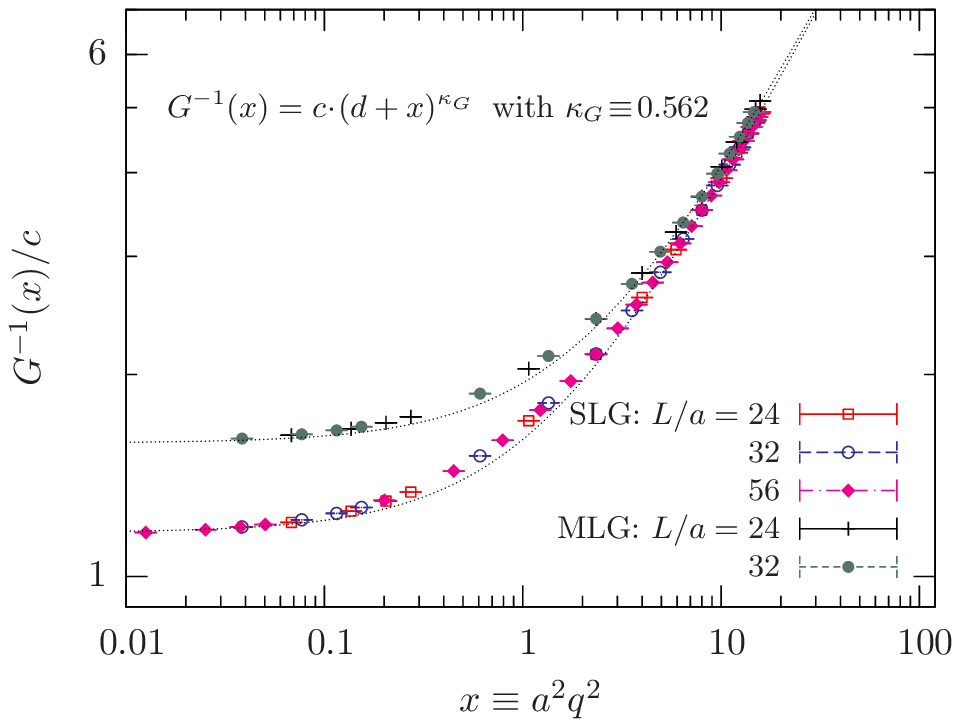}  
    \caption{Inverse ghost dressing functions in the strong-coupling
      limit of minimal lattice Landau gauge using SLG and
      MLG gauge fields/conditions. Data is normalised to the 
      scaling branch after fitting to \mbox{$G^{-1}(x)=c(d+x)^{\kappa_G}$};
      all with $\kappa_G=0.562$}
    \label{fig:ghinv_dress_qq_beta0}
  \end{minipage}\vspace{-2ex}
\end{figure*}
The data for the gluon propagator of SLG (red filled diamonds) is
compared to that of MLG (blue filled circles) in
\Fig{fig:gl_qq_beta0}. There we also show data for the gluon
propagator where either $aA^{\mathsf{adj}}_{x\mu} =u^0_{x\mu}
u^{a}_{x\mu}\sigma^a$ (no sum $\mu$), or $aA^{\mathsf{ln}}_{x\mu}=
\phi^a_{x\mu}\sigma^a/2$ from $U_{x\mu}=e^{i\phi^a_{x\mu}\sigma^a/2}$
were used to define lattice gluon fields based on the adjoint 
representation \cite{Langfeld:2001cz}, $A^{\mathsf{adj}}$ (black open
diamonds), or on the tangent space at the identity $A^{\mathsf{ln}}$
(green crosses).  In these two cases, $A^{\mathsf{adj}}$ and
$A^{\mathsf{ln}}$, for the purpose of a qualitative comparison, we
simply use the gauge configurations of the SLG to calculate the gluon
propagator.  Especially for $A^{\mathsf{ln}}$ this implies, however,
that the condition $q_{\mu}(k)A_{\mu}(k)=0$ is satisfied at best
approximately and nowhere near the precision of SLG or MLG. This
uncertainty then causes the somewhat larger errors for
this definition as seen in \Fig{fig:gl_qq_beta0}.

First, we fit the data from all four definitions to
$D(x)=c(d+x)^{2\kappa_Z-1}$ which provides the best overall description
in the full momentum range. In order to demonstrate how the other
definitions compare to the SLG, we keep its value for the exponent
fixed when fitting the other data, i.e., $\kappa_Z = 0.562$ as
obtained for $N = 32$ in SLG is used in all fits. Relative to the
scaling branch for large~$x$ we then observe a strong definition
dependence in the (transverse) gluon mass term at small~$x$ (see
\Fig{fig:gl_qq_beta0}). The relative weight of the two asymptotic
branches, scaling at large~$x$ and massive at small, is clearly
discretisation dependent and can not be compensated by finite renormalisations. 
A first indication that the massive branch might indeed be the
ambiguous one is the observed  $M\propto 1/a$. This is consistent with the
fact that the definitions of gauge fields on the lattice, which agree
at leading order, all differ at order $a^2$, and so do their
corresponding Jacobian factors which leads to lattice mass
counter-terms of different strengths. We find a similar behaviour of
the ghost dressing function whose inverse is shown in
\Fig{fig:ghinv_dress_qq_beta0}.

\section{Conclusion}

The Landau-gauge gluon and ghost propagators in the strong-coupling
limit do in fact show the scaling behaviour as predicted by the
continuum studies mentioned above. A comparison of various lattice
definitions of gauge potentials, all equivalent in the continuum
limit, shows that critical exponent and coupling can be extracted
from the high-momentum data, with $a^2q^2 > 1$, in the strong-coupling
limit in good agreement with the continuum predictions, $\kappa_Z =
\kappa_G\approx0.595$. The deviations from this scaling behaviour, on
the other hand, depend on the choice of the lattice definition of the
gluon fields, i.e., the massive branch observed for $a^2q^2 <1$.

\begin{figure*}
  \centering
  \begin{minipage}{0.49\linewidth}
    \includegraphics[width=\linewidth]{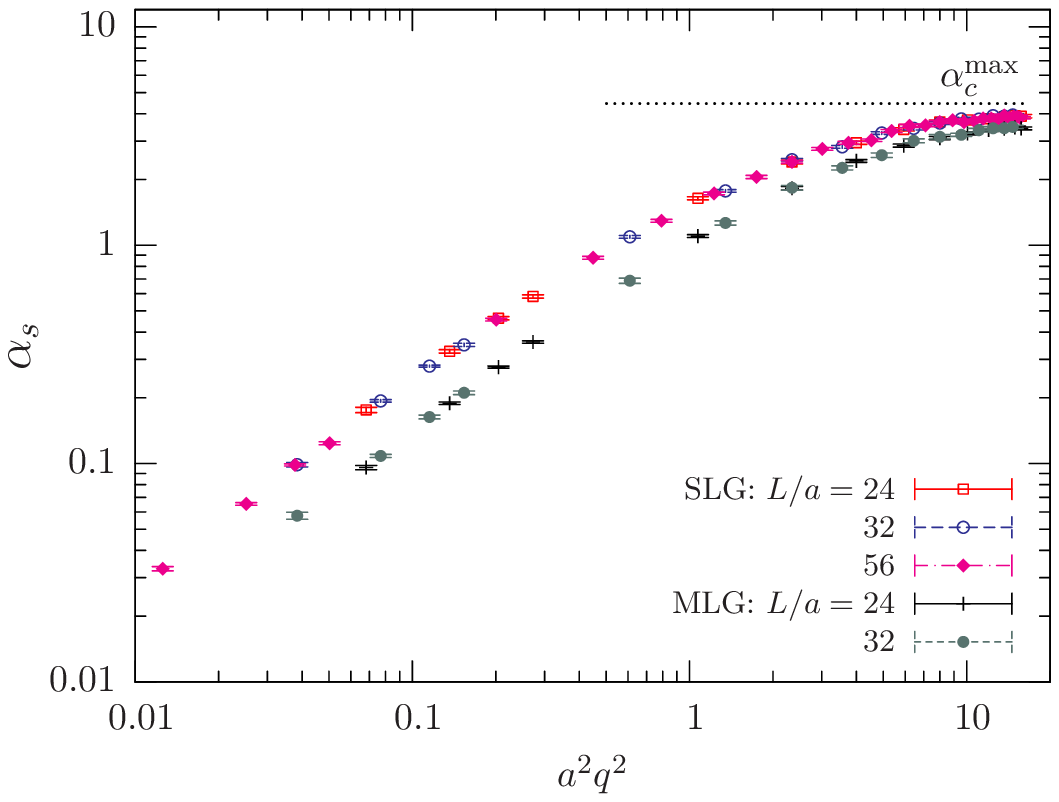}
    \vspace{-3ex}
    \caption{$\alpha_s$, for the standard (SLG)
      and modified (MLG) lattice Landau gauge at $\beta=0$. The dotted
      line is the critical coupling $\alpha^{\mathrm{max}}_c\approx4.46$
      for $N_c=2$.}
    \label{fig:alpha_qq_beta0}
  \end{minipage}
  \hfill
  \begin{minipage}{0.49\linewidth}
    \includegraphics[width=\linewidth]{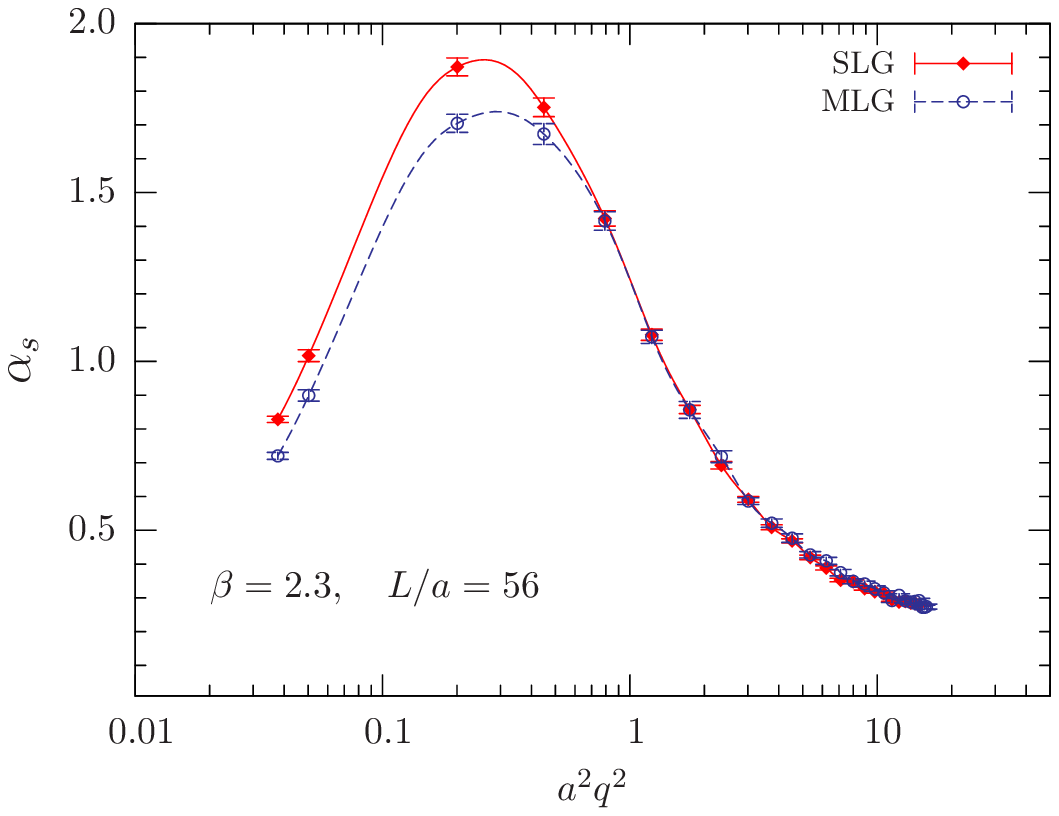}
    \vspace{-3ex}
    \caption{$\alpha_s$ for the standard (SLG) and modified
      (MLG) lattice Landau gauge at $\beta=2.3$ on a $56^4$
      lattice. Lines are spline interpolations to guide the eye.}
    \label{fig:alpha_qq_beta2p3}
  \end{minipage}\vspace{-2ex}
\end{figure*}

In complete agreement with this, the coupling (1.2) for large $a^2q^2$
levels at $\alpha_c \approx 4$, just below the upper bound
$\alpha^{\mathrm{max}}_c \approx 4.46$ for $SU(2)$, while violations
to this conformal scaling, set in as soon as the ambiguity in the
definition of minimal lattice Landau gauge does (see
\Fig{fig:alpha_qq_beta0}). Nonetheless, it is quite compelling that
the result, $\alpha_c\approx4 $, is nearly independent of the
gauge-field definition.

Indeed, the strong deviations at small momenta are linked to the strong-coupling
limit in which discretisation effects are enhanced to the
extreme and it is still possible that they disappear in the
continuum limit, eventually. But because it is a combination of
ultraviolet (mass counter-term) and infrared (breakdown of STIs)
effects, this might take very fine lattice spacings in combination
with very large volumes and therefore who-knows-how big lattices to
verify explicitly. Note that these effects are definitely persist for
commonly used $\beta$ for SU(2) (see \Fig{fig:alpha_qq_beta2p3}
where data from SLG and MLG fixed configurations at
$\beta=2.3$ are shown). These observed differences persist for $\beta
= 2.5$.

\vspace{-1ex}

\small

\begin{thebibliography}{10%
}\setlength{\itemsep}{-2pt}
\vspace{-2ex}
\bibitem{vonSmekal:1997isvonSmekal:1997vx}
L.~von Smekal, A.~Hauck and R.~Alkofer {\em Phys. Rev. Lett.} {\bf 79} (1997)
  3591; {\em Ann. Phys.} {\bf 267} (1998) 1.

\bibitem{Lerche:2002ep}
C.~Lerche and L.~von Smekal {\em Phys. Rev.} {\bf D65} (2002) 125006.

\bibitem{Zwanziger:2001kw}
D.~Zwanziger, {\em Phys. Rev.} {\bf D65} (2002) 094039.

\bibitem{Pawlowski:2003hq}
J.~M. Pawlowski, D.~F. Litim, S.~Nedelko, and L.~von Smekal {\em Phys. Rev.
  Lett.} {\bf 93} (2004) 152002.

\bibitem{Alkofer:2004it}
R.~Alkofer, C.~S. Fischer, and F.~J. Llanes-Estrada {\em Phys. Lett.} {\bf
  B611} (2005) 279--288.

\bibitem{Fischer:2006vf}
C.~S. Fischer and J.~M. Pawlowski {\em Phys. Rev.} {\bf D75} (2007) 025012.

\bibitem{Sternbeck:2007ug}
A.~Sternbeck, L.~von Smekal, D.~B. Leinweber, and A.~G. Williams {\em PoS} {\bf
  LAT2007} (2007) 340.

\bibitem{Bogolubsky:2007ud}
I.~L. Bogolubsky, E.--M. Ilgenfritz, M.~M{\"u}ller-Preussker, and A.~Sternbeck
  {\em PoS} {\bf LAT2007} (2007) 290.

\bibitem{Cucchieri:2007md}
A.~Cucchieri and T.~Mendes, {\em PoS} {\bf LAT2007} (2007) 297.

\bibitem{Fischer:2008uz}
C.~S. Fischer, A.~Maas, and J.~M. Pawlowski preprint
  \href{http://xxx.lanl.gov/abs/0810.1987}{{\tt arXiv:0810.1987}}.

\bibitem{Sternbeck:2008wgSternbeck:2008mv}
A.~Sternbeck and L.~von Smekal {\em PoS} {\bf LATTICE2008} (2008) 267; and
preprint \href{http://xxx.lanl.gov/abs/0811.4300}{{\tt arXiv:0811.4300}}.

\bibitem{vonSmekal:2007ns}
L.~von Smekal, D.~Mehta, A.~Sternbeck, and A.~G. Williams {\em PoS} {\bf
  LAT2007} (2007) 382.\\*[-0.8ex]
L.~von Smekal, A.~Jorkowski, D.~Mehta and A.~Sternbeck,
  \href{http://xxx.lanl.gov/abs/0812.2992}{{\tt arXiv:0812.2992}}.

\bibitem{Langfeld:2001cz}
K.~Langfeld, H.~Reinhardt, and J.~Gattnar {\em Nucl. Phys.} {\bf B621} (2002)
  131--156.
\end{thebibliography}

\providecommand{\href}[2]{#2}\begingroup\raggedright\endgroup

\end{document}